\documentclass[final,3p,times,twocolumn]{elsarticle}
\usepackage{epsfig}

\biboptions{}

\journal{Physics Letters A}

\begin{document}

\begin{frontmatter}


%\title{On a possibility to calculate
%the fine structure constant
%in the compensation approach}
%\author{B.A. Arbuzov and I.V. Zaitsev}
%\ead{arbuzov@theory.sinp.msu.ru}
%\address{M.V. Lomonosov Moscow State University,  119991 Moscow, Russia}

%\title{Associated heavy particles production with Higgs  as a tool for
%a search for non-perturbative effects of the electroweak interaction
%at the LHC}
\title{Non-Lagrangian equations of motion and the momentum non-conservation}
\author[a]{B.A. Arbuzov}
\ead{arbuzov@theory.sinp.msu.ru}
\address[a]{M.V. Lomonosov Moscow State University,
 119991 Moscow, Russia}

%\author{}

%\address{}

\begin{abstract}
In the framework of a geometrical model, in which the affine connection of a space is expressed in terms of the electromagnetic field, a possibility of the momentum non-conservation is shown. A
toy device with an object moving in a magnetic field is presented for an illustration of
the effect.
\end{abstract}


\end{frontmatter}

% \linenumbers
\section{Introduction}\label{sec1}

The basic principle of the energy and the momentum conservation is considered to form the foundation of the physics. The more so some  surprising experimental results need efforts for the proper interpretation.

In recent work~\cite{NASA} the authors claim for the following effect for thrust to power ratio $R$
\begin{equation}
R\,=\,(1.2 \pm 0.1)\,\frac{mN}{kW}\,;\label{NASA}
\end{equation}
for a closed radio-frequency cavity in vacuum with the resonance frequency $1.937\,GHz$, which is loaded by a dielectric disk.

Such results cause natural distrust due to the evident contradiction to the momentum conservation. However it may be advisable to consider the question: what to do provided these dubious results nevertheless will be confirmed?

We are to note, that a possibility of the energy and momentum non-conservation was discussed in few old works~\cite{AF67,Arb68,Arb72}, connected with a study of an affine geometry in an application
to a model of a geometrical interpretation of the $CP$-violation.

These works were initiated
by the substantiation of the $CP$ symmetry in terms of properties of the space-time.
Let us recall works by L.D. Landau~\cite{LDL} and E.P. Wigner~\cite{Wigner}, in which
the sound arguments were expressed on behalf of the validity of the $CP$ symmetry after
the discovery of the parity non-conservation. The main point here is that in the Minkowski
space operators of the time shift and the space reflection commute, therefore we
are forced to admit that there is some conservation law for an operator, which corresponds to  the space reflection. With parity $P$ being evidently violated, we have to multiply
$P$ by some universal operator, which is not straightly connected with space-time variables.
At that time, as well as nowadays, the only candidate for this part was just charge conjugation $C$. After the discovery of the $CP$-violation the argumentation of
works~\cite{LDL,Wigner} somehow lost an actuality.

However there were also attempts to consider the actual $CP$-violation without contradiction to
arguments~\cite{LDL,Wigner}. Then naturally one has to consider a generalisation of
the Minkowski space. In particular, a variant of a modification of properties of the space was considered in work~\cite{AF67}  in application to a model of $CP$-violation in which the modification
was connected with the antisymmetric tensor $E_{\mu\,\nu}$ of the electromagnetic field.
Note, that attempts to include electromagnetic field into geometrical schemes were repeatedly undertaken. Indeed, the gravitational and the electromagnetic fields are the only long-range  ones known, and their influence might become apparent in properties of the space.

In work~\cite{Arb68} the expression for an affine connection of the space
in presence of both the gravitational and the electromagnetic fields was proposed.
In the framework of the geometrical model~\cite{AF67,Arb68, Arb72} in addition to interpretation of the possibility of the $CP$-violation there could
appear other effects, which need a special consideration. First of all, we are to consider the problem of conservation laws. It comes out, that unusual effects could arise due to a special form of equations of motion.

We are acquainted with two ways to obtain equations of motion. The first one
consists in a formulation of a Lagrangian of the system under consideration and then Euler
equations for this Lagrangian are just the equations of motion. In case the Lagrangian is translation invariant in respect to time and space shifts, there are conservation laws for
the energy and the momentum. Due to the space-time uniformity the energy-momentum conservation   is considered to be firmly established.

However there is another way to obtain equations of motion. In the General Relativity
a test body moves along a geodesic line, which is defined by the affine connection $L_{\mu \nu}^\alpha$. The equations of a geodesic line, which really are equations of motion are the following
\begin{equation}
\frac{d^2\,x^\mu}{d\,s^2}
\,+\,L^\mu_{\rho\,\sigma}\,
\frac{d x^\rho}{d s}\,\frac{d x^\sigma}{d s}
\,=\,0\,;\label{geodesic}
\end{equation}
where $s$ is a proper time.
In the General Relativity the affine connection coincides with Christoffel symbols
$L^\mu_{\rho\,\sigma} =\Gamma^\mu_{\rho\,\sigma}$, which are symmetric in respect to
lower indices $\rho\,\sigma$ and we obtain usual conservation laws.

Let us emphasize, that in case of a general affine space, affine connection $L_{\mu \nu}^\alpha$ is not obligatory
symmetric. For example, for spaces with absolute parallelism~\cite{Cartan, Ein30} there is also antisymmetric part of the affine connection, which leads to a nonzero tensor of torsion
\begin{equation}
S^\mu_{\rho\,\sigma}\,=\,\frac{1}{2}\,\bigl(L^\mu_{\rho\,\sigma}-L^\mu_{\,\sigma \rho}\bigr).
\label{torsion}
\end{equation}
In case of a nonsymmetric affine connection the equation for geodesic line~(\ref{geodesic}) is not corresponding to some Lagrangian and so there is no obligatory
conservation laws.

In work~\cite{Arb72} wouldbe effects in such space were discussed. Let us show the affine
connection $L_{\mu \nu}^\alpha$  for a space with the absolute parallelism~\cite{AF67, Arb72}, in case of the affine connection being defined by the electromagnetic field
\begin{eqnarray}
& &L_{\mu \nu}^\alpha=(\delta\,^\alpha_\rho + F^{\,\alpha \cdot}_{\cdot\rho})\,\partial_\nu F^{\,\cdot \rho}_
{\mu\, \cdot}\,;\label{L}\\
& & F^{\,\alpha \,\cdot}_{\cdot\, \rho} = g^{\,\alpha \,
 \sigma}F_{\sigma \rho}\,,\quad F^{\,\cdot\, \rho}_{\mu\, \cdot} = g^{\,\rho\, \sigma}
F_{\mu \,\sigma}\,;\nonumber
\end{eqnarray}
where $F_{\mu \nu}$ is a tensor, satisfying the following condition
\begin{equation}
(\delta^\alpha_\sigma + F^{\alpha\, \cdot}_{\cdot\, \sigma})(\delta^\sigma_\nu + F^{\,\cdot \,\sigma}_{\nu\, \cdot}) = \delta^\alpha_\nu\,. \label{condition}
\end{equation}
This relation defines symmetric part of tensor $F_{\mu \nu}$ in terms of its antisymmetric part, which
can be related to the antisymmetric tensor of the electromagnetic field $E_{\mu\,\nu}$ in the following manner~\cite{AF67}
\begin{equation}
\frac{e}{2}\,(F_{\mu\;\nu}-F_{\nu\,\mu}) = \pm\, l_0^2\,E_{\mu\,\nu};
\label{Emn}
\end{equation}
where $l_0$ is a new constant of the length dimension, which was introduced in
work~\cite{AF67} and $e$ is the elementary electric charge $e=1.602\cdot10^{-19}\,C$. In works~\cite{AF67}-\cite{Arb72} two suitable scales of parameter $l_0$ are discussed
\begin{equation}
l_{0\,1} \simeq 10^{-14}\,cm;\quad l_{0\,2} \simeq 10^{-17}\,cm.\label{l0}
\end{equation}
Let us note, that the sign in relation~(\ref{Emn}) is to be defined by further studies, provided they will be performed.

Relations~(\ref{condition}, \ref{Emn}) define tensor $F_{\mu \nu}$ in terms of electromagnetic field tensor $E_{\mu\,\nu}$.

Affine connection~(\ref{L}) defines equations of motion, which according to rules of a geometry
are just equations of a geodesic line~(\ref{geodesic}).

It is very important, that in case the antisymmetric part of $F_{\mu\,\nu}$ being proportional to  $E_{\mu\,\nu}$ is not zero, equations~(\ref{geodesic}) are non-Lagrangian ones~\cite{Arb68,Arb72}, that is they can not be referred to some Lagrangian. Therefore conservation lows are not valid for a motion, which is
described by equations~(\ref{geodesic}). Applications for wouldbe energy non-conservation are described in work~\cite{Arb72} and estimate~(\ref{l0}) of parameter $l_0$ is obtained according to absence of contradictions in the data for solar luminosity. The possibility of the momentum non-conservation was also noted in work~\cite{Arb72}.

Let us write down the expression for the non-relativistic force in the first order of
$l_0^2$, which follows from
equation~(\ref{geodesic}) with account of relations~(\ref{L}, \ref{condition}, \ref{Emn})
\begin{eqnarray}
& &\overline{F}\, \,= \mp\, \frac{m\, c\, l_0^2}{e}\,\Bigl(\,\frac{\partial \overline{E}}{\partial t}\,+\,(\overline{v}  \,\bigtriangledown) \overline{E}\,+\nonumber\\
& &\frac{\overline{v}}{c}\times\frac{\partial\overline{H}}{\partial t}\,+\,
\frac{\overline{v}}{c}\times (\overline{v}\, \bigtriangledown)\overline{H}\,\Bigr)\,;\label{force}
\end{eqnarray}
where notation $\overline{a}$ means spatial vector $a_k$, $\overline{E}$ and $\overline{H}$ are, as usually,
an electric and a magnetic fields and $\overline{v}$ is a velocity of a moving object.
\section{A toy model for a device with the momentum non-conservation}
\label{sec2}
\begin{figure}
\begin{center}
\includegraphics[scale=0.45]{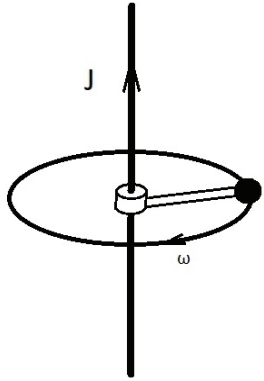}
\caption {The scheme for a device for the illustration of the momentum non-conservation with an action of force~(\ref{force}).}
\label{fig:Oneloop}
\end{center}
\end{figure}
Let us consider the device, consisting of some dielectric ball with mass $m$, which rotates
with angular velocity $\omega$ around a
vertical wire with current $J$, as it is presented in Fig.~\ref{fig:Oneloop}.
Current $J$ generates magnetic field
\begin{equation}
H\,=\,\frac{2 J}{c\,R}\,;\label{H}
\end{equation}
which is directed along the trajectory of the ball. The result for force $\overline{F}$ is given by the fourth term in expression~(\ref{force}), while other three terms do not contribute. Now we have with $x,\,y$ being coordinates of the ball in the plane of the rotation
\begin{eqnarray}
& &H_x=-\frac{2\,J\,y}{c\,R^2},\;H_y=\frac{2\,J\,x}{c\,R^2};\nonumber\\
& &R=\sqrt{x^2+y^2},\,
v_x=-\omega\,y,\;v_y=\omega\,x\,;\nonumber\\
& & U_x = v_x \frac{\partial}{\partial x}H_x+v_y \frac{\partial}{\partial y}H_x =
 -\frac{2 J\, \omega\, x}{c R^2}\,;\label{Force}\\
& & U_y = v_x \frac{\partial}{\partial x}H_y+v_y \frac{\partial}{\partial y}H_y =
 -\frac{2 J\, \omega\, y}{c R^2};\nonumber\\
& &v_x\,U_y-v_y\,U_x\,=\,-\frac{2\,J\,\omega^2}{c}.\nonumber
\end{eqnarray}
From the last line in~(\ref{Force}) we see, that the vector product in expression~(\ref{force}) is directed vertically and we obtain the following vertical force
\begin{equation}
F_z = \mp\, \frac{2\, m\, l_0^2\, J\, \omega^2}{e \,c}\,.\label{forcew}
\end{equation}
This force leads to an appearance of the vertical thrust due only to inner processes in the system.
Thus we have a simple example of the momentum non-conservation with a non-Lagrangian
equation of motion.

\section{Conclusion}
The example of the possibility of the momentum nonconservation illustrates
a wouldbe situation with a non-Lagrangian equation of motion. An application
of analogous considerations to results of work~\cite{NASA}
needs further special studies. For the moment we can not affirm, that equations~(\ref{geodesic}) with account of~(\ref{L},\ref{condition},\ref{Emn}) really
explain results~\cite{NASA}. Note, that in both cases, that is with our example and with device~\cite{NASA}, there are dielectric samples, which are to be pushed by
a new force. However in our case the sample is moving in the magnetic field provided by
current $J$, and in~\cite{NASA} it is motionless, but
an electric and a magnetic fields are alternating. In this case $\bar v=0$ we have the following
expression of the force for the first two terms in expansion in powers of $(l^2_0/e)$
\begin{equation}
\overline{F}\, = \mp \frac{m  c  l_0^2}{e}\,\Bigl[\frac{\partial\overline{E}}{\partial t}\mp
\frac{l_0^2}{2\,e}\Bigl(\overline{H}\times\frac{\partial \overline{E}}{\partial t}-\frac{\partial \overline{H}}{\partial t}\times\overline{E}\Bigr)\Bigr].\label{v0}
\end{equation}
The second term in square brackets of~(\ref{v0}) is a non-Lagrangian one and thus effect of
the momentum non-conservation is proportional to $l_0^4/e^2$. Results for
an action of force~(\ref{v0}) under conditions being discussed in work~\cite{NASA} need
special detailed calculations. The first remark, which we may express here concerning comparison with
results~\cite{NASA}, consists in the prediction, that the wouldbe effect for force~(\ref{v0}) is to be proportional to the power, because
 the energy density of the electromagnetic field is proportional to the fields squared  $(E^2+H^2)/8 \pi$. The data, presented in work~\cite{NASA}, agree with such dependence.

Let us also roughly estimate  contribution of force~(\ref{v0}) to a device analogous to that presented
in~\cite{NASA}. Considering properties of a stationary electromagnetic wave with orthogonal
$E$ and $H$ shifted in phases by $\pi/2$ we have from~(\ref{v0}) the folowing estimate for the force
\begin{equation}
F\,\simeq\,\frac{m c l_0^4 \omega E^2}{2\,e^2}\,\simeq\,\frac{4 \pi^2\,\rho\,\nu_R\, Q\,W\,l_0^4}{e^2};
\label{FF}
\end{equation}
where $\rho$ is the density of the sample, $\omega = 2\,\pi\,\nu_R $ is the resonance circular frequency, $W$ is the power
applied to the device and $Q$ is a quality factor.

Thus ratio R~(\ref{NASA}) could be estimated as follows
\begin{equation}
R\,=\,\frac{F}{W}\,\simeq\,\frac{4 \pi^2\,c\,\rho\,\nu_R\, Q\,l_0^4}{e^2}.\label{F/W}
\end{equation}
From~(\ref{F/W}) we might get estimation of $l_0$, which could lead to result~(\ref{NASA}).
For example with values $\nu_R=1.937 GHz ,\,  \rho=1.0\,g/cm^3,\, Q=10^6$ and with wouldbe result~(\ref{NASA}) we have
\begin{equation}
l_0\,\simeq\,3.4\cdot10^{-14}\,cm\,;\label{L0}
\end{equation}
that may be confronted with estimates~(\ref{l0}). Result~(\ref{L0}) seems to be consisent with the first possibility in~(\ref{l0}).

In any case the examples of non-standard forces, which we have discussed here, might be instructive provided
effects of the momentum non-conservation similar to that being presented in~\cite{NASA} would persist.
%\newpage

\section{Acknowledgments}
The work is supported in part by grant NSh-7989.2016.2 of the
President of Russian Federation.

The author expresses a gratitude to V.I. Savrin and I.V. Zaitsev for fruitful
discussions and the helpful assistance.

\end{document}